# Mechanical Metastructures of Triple Periodic Carbon Clathrates

Jinghui Zhu, Ke Xu, Zhisen Zhang[*], Jianyang Wu[*]

Department of Physics, Research Institute for Biomimetics and Soft Matter, Jiujiang Research Institute and Fujian Provincial Key Laboratory for Soft Functional Materials Research, Xiamen University, Xiamen 361005, PR China

**Abstract:** Clathrates are lightweight, cage-like, fully-$sp^3$ three dimensional (3D) structures that are experimentally-available for several host elements of the IV group. However, carbon clathrates are as yet hypothetical structures. Herein, the mechanical properties of Type-I-$C_{46}$ Type-II-$C_{34}$ and Type-H-$C_{34}$ carbon clathrates are explored by first-principles calculations. It is revealed that those carbon clathrates show distinct anisotropic patterns in ideal tensile/shear strengths and critical tensile/shear strains, with maximum ideal tensile strength of Type-I carbon clathrate that is superior over that of diamond in <111> direction. However, it is identified isotropy in shear Young's modulus, and in terms of tensile/shear Young's moduli, they are sorted as Type-I > Type-II > Type-H carbon clathrates. There are distinct critical load-bearing bond configurations that explain their distinct mechanical behaviors. Moreover, those carbon clathrates are intrinsically indirect semiconductors, and their electronic properties can be greatly dictated by mechanical strain. Carbon clathrates can be potentially utilized as lightweight technically robust engineering metastructures and in electromechanical devices.

[*]Corresponding Emails: zhangzs@xmu.edu.cn, jianyang@xmu.edu.cn

# 1. Introduction

Carbon is one of the most well-known and important elements in the periodic table. In natural settings, there are mainly three carbon allotropes including diamond, graphite and amorphous carbon-based structures. Among them, diamond is the most mechanically robust substance in nature. Due to its super high strength and hardness, diamond plays an irreplaceable role in the field of mechanical processing[1]. In recent decades, many other carbon allotropes have been discovered/synthesized in laboratory-settings[1-7]. For example, in 1985, Kroto et al. observed a newly carbon cluster structure molecule consisting of 60 carbon atoms, named as buckminsterfullerene[2]. In 1991, Iijima et al. discovered carbon nanotubes (CNTs) and opened up a new field of researching for one-dimensional (1D) nanomaterials[3]. In 2004, two-dimensional (2D) graphene was first mechanically stripped from graphite by Novoselov et al.[6] Those carbon allotropes show excellent mechanical properties and electrical properties due to their unique $sp^2$-hybridized C-C bonds[4, 5, 7].

Theoretically, a number of carbon structures have been also predicted/proposed[8-19]. For example, Gang et al.[19] predicted a new three-dimensional (3D) carbon cubic crystal structure composed of $sp^3$-hybridized C-C bonds, named as T-carbon, which is structurally characterized by that atoms in the cubic diamond are replaced by carbon tetrahedrons. Soon afterwards, this unique T-carbon structure was successfully synthesized by picosecond laser irradiation of multi-walled carbon nanotubes (MWCNTs) suspended in methanol solution[14]. Benedek et al. proposed three lattice structures of carbon clathrates with 3D $sp^3$-hybridized C-C bonded networks, named as *fcc*-$C_{136}$, *sc*-$C_{46}$ and *hex*-$C_{40}$, respectively[11]. Note that similar IV-group elemental (for example, Si and Ge) clathrate structures have been experimentally synthesized[8-10, 12, 13, 15-18], but to date not for those carbon clathrates.

Meanwhile, there have been a number of studies focused on their stability, elastic, and electronic properties, and so on[20-30]. For example, Wang et al.[26] calculated the structural stability of carbon clathrates at high pressure and identified *fcc*-$C_{136}$ clathrate as the third most stable carbon phase after cubic diamond and hexagonal graphite. A pressure-induced phase transition was predicted to occur around 17 GPa from hexagonal graphite to *fcc*-$C_{136}$, which is more stable than other carbon clathrates. Blase et al.[21] studied the ideal tensile and shear strengths of $C_{46}$ carbon clathrates by ab initio calculations. It was revealed that its bulk modulus and elastic constants are less than diamond, but its strengths in all directions are larger than that of the critical stresses associated with the diamond {111} planes of easy slip. Moreover, the quasi-particle band structures of Type-I and Type-II carbon clathrates have been also investigated[20]. It was found that the quasi-particle correction is similar to the corresponding diamond phase and shows a near-direct band gap of 5.15 eV.

Interestingly, triple periodic carbon clathrates are able to show unique properties comparable to those of diamond. From structural point of view, a large variety of triple periodic carbon clathrates can be designed by adjusting the basic building blocks of polyhedral cages in clathrate structures. In particular, as a result of unique $sp^3$-hybridized C-C bonded clathrate cages, triple periodic carbon clathrates are expected to be lightweight, mechanically and electrically robust metastructures, enabling them as practical mechanical engineering materials. In this work, tensile and shear mechanical characteristics of three distinct triple periodic carbon clathrates (Type-I, Type-II and Type-H) that are made up of a variety of polyhedral cages are comprehensively explored, as well as their mechanical strain-dependent electronic properties, using first-principle calculations.

## 2. Models and Methodology

### 2.1 Structural Models of Carbon Clathrates

Inspired by structural I, II and H clathrate hydrates, triple periodic Type-I, -II and -H carbon clathrates are constructed and taken into investigations. From molecular point of view, as shown in Figure 1, triple periodic Type-I and Type-II carbon clathrates are mainly designed by well-stacking two basic structures of $5^{12}+5^{12}6^2$ and $5^{12}+5^{12}6^4$ polyhedral cages, respectively, while Type-H carbon clathrate is structurally dominated by $5^{12}+4^35^66^3+5^{12}6^8$ polyhedral cages. The primary unit-cell of those three clathrate structures contains 46, 34, and 34 carbon atoms, respectively. Because the C-C bond length in triple periodic carbon clathrate is much shorter than that of hydrogen (H-) bonds in clathrate hydrates, there is precious little room in polyhedral cages of triple periodic carbon clathrates. Therefore, unlike other clathrate structures, triple periodic carbon clathrates are guest-atom/molecule-free structures[28].

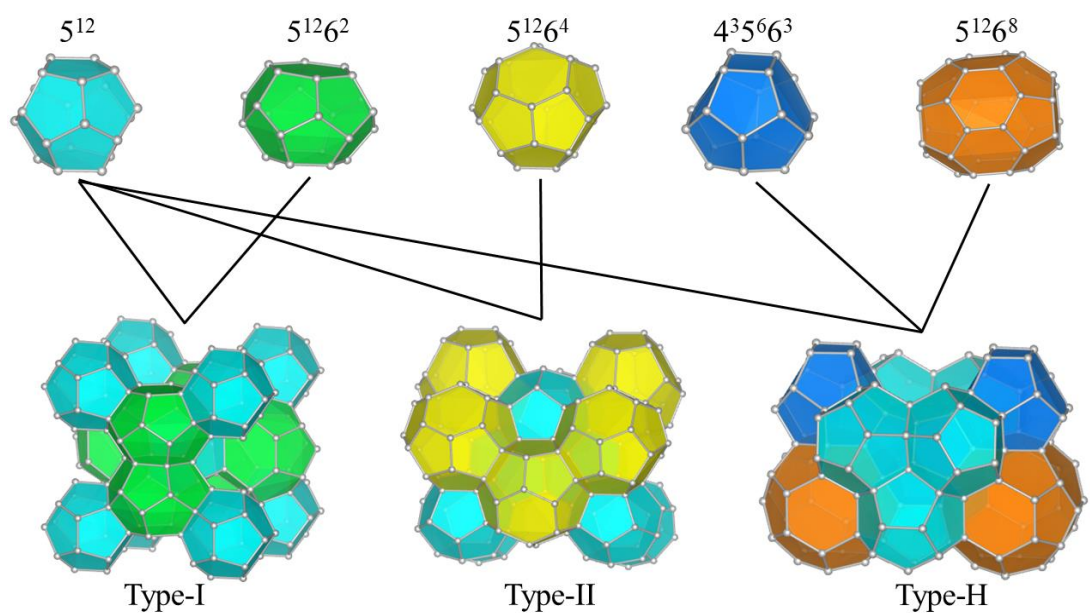

Figure 1 Molecular structures of triple periodic clathrate hydrates. Triple periodic Type-I-$C_{46}$, Type-II-$C_{34}$ and Type-H-$C_{34}$ carbon clathrates are structurally dominated by a combination of $5^{12}+5^{12}6^2$, $5^{12}+5^{12}6^4$ and $5^{12}+4^35^66^3+5^{12}6^8$ polyhedral cages, respectively, where the polyhedral cages are colored for clarification.

## 2.2 Methodology

Mechanical properties of those three triple periodic carbon clathrates are examined by means of first-principle calculations in the framework of density functional theory (DFT) as implemented in the Vienna ab initio Simulation Package (VASP)[31, 32]. The projector augmented wave (PAW) method[33] is employed for interactions between ion cores and valence electrons. The electron exchange-correlation interactions are described by the generalized gradient approximation (GGA) in the form proposed by Perdew-Burke-Ernzerhof (PBE)[34]. The plane-wave cutoff energy is assigned as 550 eV expansion of valence electron wave functions. The first Brillouin zone is sampled by $\Gamma$-centered $2 \times 2 \times 2$, $3 \times 3 \times 3$ and $3 \times 3 \times 3$ $k$-grid mesh generated by the Monkhorst-Pack scheme for Type-I, -II and -H carbon clathrates, respectively. The relaxation of triple periodic carbon clathrates considering both atomic positions and lattice vectors is performed until the total energy is converged to $1.0 \times 10^{-5}$ eV/atom and the maximum force on each atom is less than 0.01 eV/Å. With regard to the calculations of band structures, the $k$-point paths and grid densities for three different clathrate structures are selected on the basis of previous studies by Setyawan et al.[35], in which the $k$-point paths of the three structures that belong to cubic, triclinic and monoclinic systems are assigned as $\Gamma$-X-S-Y-$\Gamma$-Z-U-R-T-Z, X-$\Gamma$-Y and $\Gamma$-M-K-$\Gamma$-A-L-H-A, respectively.

## 3. Results and Discussion

### 3.1 Mechanical Responses of Carbon Clathrates

Figure 2 shows the resulting uniaxial tensile and shear mechanical responses of triple periodic Type-I, -II and -H carbon clathrates subjected to specific crystalline loading directions as illustrated by the insets, as well as the mechanical responses of mechanically robust diamond from previous studies[36]. Apparently, similar to diamond, all investigated carbon clathrates exhibit three deformational stages on the basis of their mechanical tensile and shear loading curves. The deformational stage I is described by the initial linear mechanical responses that correspond to the linear elastic behavior. Both tensile and shear Young's moduli can be obtained by linearly fitting the linear stress-strain curves. In terms of tensile/shear Young's modulus, they are sorted as: Type-I (870.9/436.3 GPa) > Type-II (799.6/414.3 GPa) > Type-H (732.6/378.8 GPa) carbon clathrates. By comparison, those carbon clathrates are less tensile stiff than diamond (1144.3 GPa), whereas in terms of shear stiffness, they are comparable to diamond[36]. The second deformational stage corresponds to the smoothly nonlinear mechanical responses up to sudden deep drops of loading stresses, and they are characterized by strain-softening behavior, namely, the slopes of mechanical stress-strain curves are gradually reduced with increasing strain. Intriguingly, it is identified negative slope within finite strain regime of around 0.20-0.26, or negative tensile stiffness for Type-I carbon clathrate, suggesting its distinct deformation mechanism from the cases of Type-II and -Hones. Upon shear load, however, it is not observed negative shear stiffness for all carbon clathrates. The ideal tensile strengths of Type-I, -II and -H carbon clathrates, defined by the upper limit of material tensile strength, are found to be approximately 100.3 GPa, 103.1 GPa and 94.9 GPa, respectively, which are higher than that of diamond (around 90.0 GPa) in the <111> loading direction. Under shearing, however, the ideal shear strength can be slightly either higher or lower than that of diamond in the {111}<112> directional shear load. In terms of idea shear strength, they are ranked as: Type-I > Type-II > Type-H carbon clathrates, differing from the case in terms of idea tensile strength. It can be summarized that those Type-I, -II and -H carbon clathrates are mechanical metastructures. The last deformational stage is primarily characterized by sudden deep drops of loading stress, indicating occurrence of failure in carbon clathrates. Upon tension, both Type-I and -II carbon clathrates show sudden drops of loading stresses to zero in this deformational stage, whereas for Type-H carbon clathrate, the loading stress drops to non-zero and almost remains constant in the following long strain regime, indicating its significant stretching plasticity. Subjected to shearing, there is a rising in the shear stresses posterior to the sudden deep drops of loading stresses, also implying remarkable shear plasticity, in sharp contrast to the case of diamond that shows non-recovery in loading shear stress posterior to the drop of stress to zero.

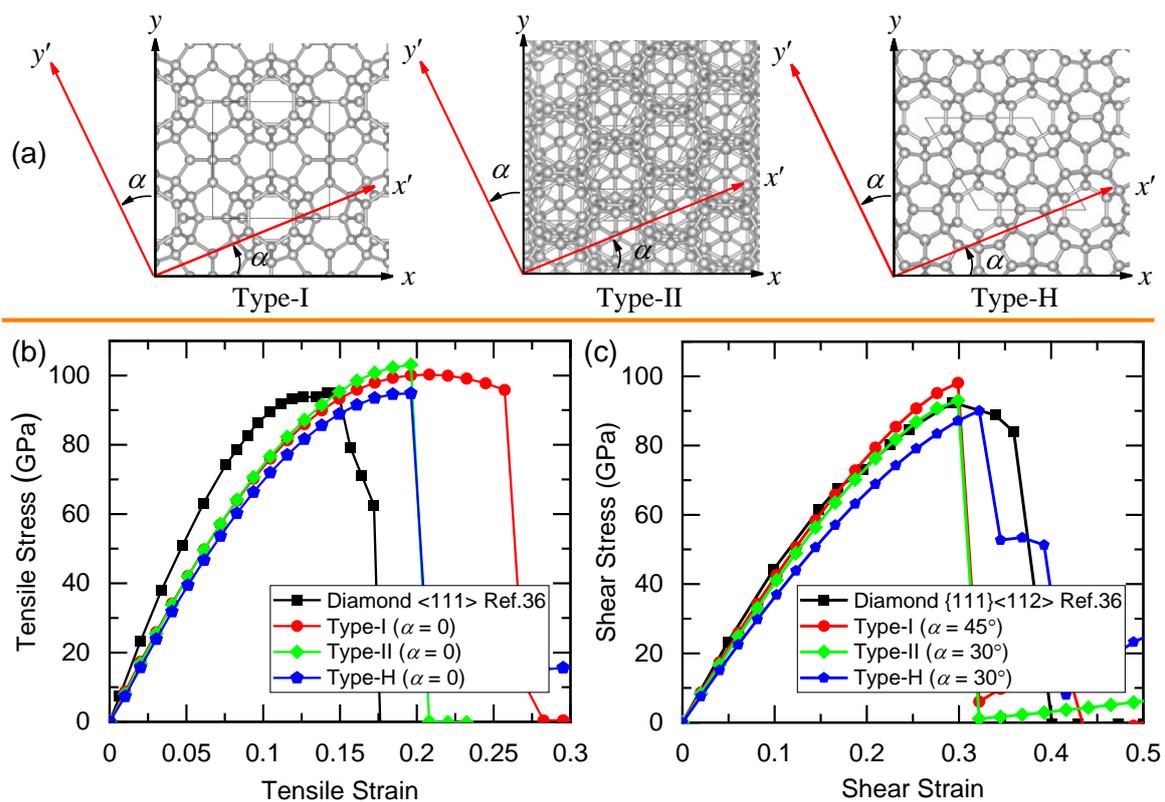

Figure 2 Mechanical responses of triple periodic carbon clathrates. (a) Illustrations of loading direction of Type-I, -II and -H carbon clathrates structures. (b) Uniaxial tensile stress-strain curves of Type-I, -II and -H carbon clathrate structures subjected to loading directions of angle $\alpha = 0°$, as well as that of bulk diamond under <111> directional uniaxial load [36] for comparison. (c) Shear stress-shear strain curves of Type-I, -II and -H carbon clathrates under loading directions of angle $\alpha = 45°$, $30°$ and $30°$, respectively, as well as that of bulk diamond subjected to the {111}<112> directional shearing for comparison.

## 3.2 Orientation-dependent Mechanical Properties of Carbon Clathrates

Due to their unique arrangement of polyhedral cages, the effects of crystalline orientation on the mechanical responses and mechanical properties of triple periodic Type-I, -II and -H carbon clathrates are also explored. As illustrated by Figure 2a, both tension and shear directions are determined by angle $\alpha$ that is defined by anti-clockwise rotation around $z$ axis. Here, the angle $\alpha$ is taken as 0°, 5°, 10°, ⋯, with an increment of 5°, and as a result of different symmetries of the three structures, the loading direction of angle $\alpha$ varies from 0 - 45°, 0 - 60° and 0 - 60° for Type-I, -II and -H carbon clathrates, respectively. Figures 3a-c show the basic mechanical properties including tensile strength, tensile Young's modulus and critical tensile strain of Type-I, -II and -H carbon clathrates subjected to different loading directions, respectively. As shown in Figure 3a, there is strong anisotropy in the tensile strength for the three carbon clathrates. Both Type-II and -H carbon clathrates exhibit similar star-like anisotropic tensile strength, in which the maximum and minimum values are subjected to loading directions of angle $\alpha$ = 0°, 60°, 120°, 180°, 240°, 300°, and $\alpha$ = 30°, 90°, 150°, 210°, 270°, 330°, respectively. By comparison, the tensile strengths of Type-II carbon clathrate are around 8.0-9.0% higher than that of Type-H one, depending on the stretching direction. However, there is distinct anisotropic pattern in the tensile strength of Type-I carbon clathrate that varies from around 93.9-100.3 GPa. For example, there are twelve peak values of tensile strength in the polar plot, within the maximum value as the loading directions are $\alpha$ = 0°, 90°, 180°, and 270°. As seen in Figure 3b, it is identified strong anisotropy in the tensile Young's modulus of Type-I carbon clathrate, within the maximum and minimum values as $\alpha$ = 45°, 135°, 225°, 315°, and $\alpha$ = 0°, 180°, respectively. However, both Type-II and -H ones show almost isotropic tension Young's modulus. For each specific loading direction, in terms of the value of tensile Young's modulus, they are ranked as Type-I > Type-II >

Type-H carbon clathrates. As shown in Figure 3c, it is observed irregular anisotropy in the critical tensile strain for all carbon clathrates, within a number of singularities in the polar plots. Moreover, the critical tensile strains are much higher than that of mechanically robust diamond.

Figures 3d-f present the ideal shear strength, shear Young's modulus and critical shear strain of triple periodic Type-I, -II and -H carbon clathrates under different loading directions, respectively. As shown in Figure 3d, there are very different anisotropy in the idea shear strength for Type-I, -II and -H carbon clathrates. For example, Type-I and -II carbon clathrates show square- and hexagon-shaped patterns of anisotropy in the ideal shear strength, whereas for case of Type-H one, it shows flower-like pattern. For both Type-II and -H carbon clathrates, the corresponding maximum and minimum ideal shear strengths are subjected to loading directions of angle $\alpha = 30°, 90°, 150°, 210°, 270°, 330°$, and $\alpha = 0°, 60°, 120°, 180°, 240°, 300°$, respectively, which is opposite to the case of ideal tensile strength. With regard to Type-I carbon clathrate, the maximum and minimum ideal shear strengths are obtained under loading directions of angle $\alpha = 45°, 135°, 225°, 315°$, and $\alpha = 0°, 90°, 180°, 270°$, respectively. Under different loading directions, they can be differently ranked in terms of ideal shear strength. Interestingly, there is negligible dependence of shear Young's modulus on the loading direction for the three carbon clathrates, although they are composed of different combination of polyhedral cages that are differently stacked. In terms of the value of shear Young's modulus obtained from any loading directions, they are sorted as: Type-I > Type-II > Type-H carbon clathrates. As shown in Figure 3f, it is identified strong anisotropic critical shear strain for the three carbon clathrates, with a characteristic of gear-like pattern. The loading directions of angles $\alpha$ for the maximum and minimum values of critical shear strain are similar to the case of ideal shear strength.

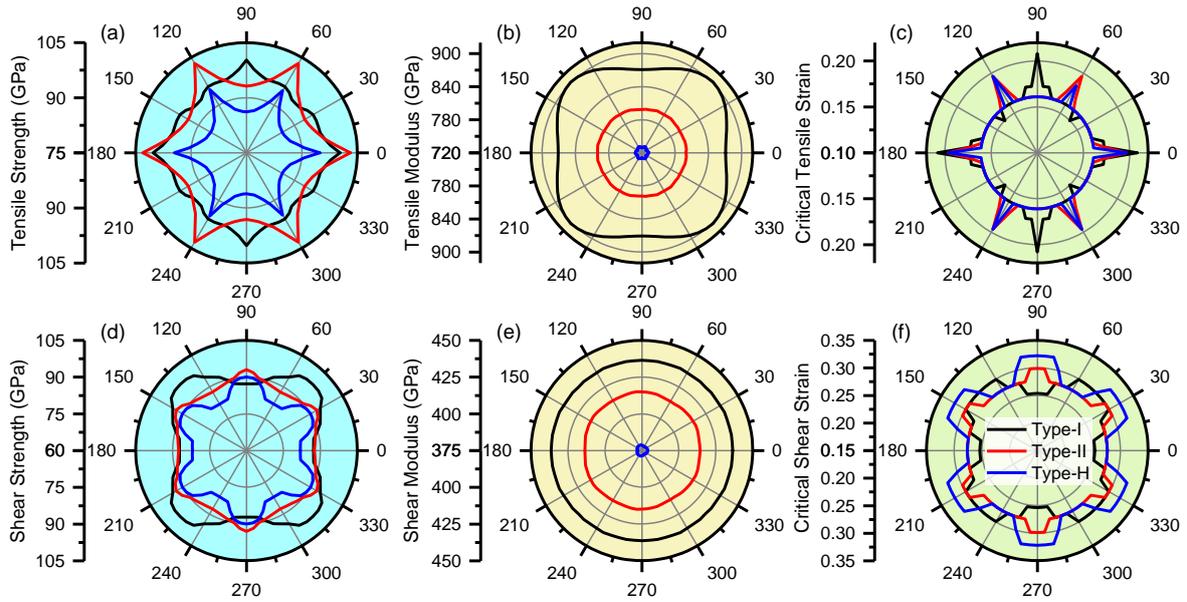

Figure 3 Crystallographic orientation dependence of (a) ideal uniaxial tensile strength, (b) tensile Young's modulus, (c) critical tensile strain, (d) ideal shear strength, (e) shear Young's modulus and (f) critical shear strain for triple periodic Type-I, -II and -H carbon clathrate structures, respectively.

**3.3 Type-I, -II and -H Carbon Clathrates as Lightweight Mechanically Robust Materials**

As illustrated by Figure 1, as-investigated Type-I, -II and -H carbon clathrates are structurally characterized by a variety of polyhedral cages, indicating that there are pretty accessible room in those structures. The mass densities of triple periodic Type-I, -II and -H carbon clathrates are calculated to be around 3.05 g/cm$^3$, 3.03 g/cm$^3$ and 2.97 g/cm$^3$, respectively, which are quite lower than that of diamond (3.50 g/cm$^3$). Figure 4 shows the tensile mechanical strength-mass density relations of a variety of crystalline carbon allotropes. Obviously, there is no apparent positive relationship between tensile strength and mass density, against the usual thinking that mechanical strength is enhanced as the mass density of materials is increased. As is seen, our studied carbon clathrates show better mechanical performance of tensile strength than that of lightweight T-carbon and heavyweight bct-C4, hP3, tI12, oS16[37], T-II carbon[38] allotropes, as well as comparable to heavyweight diamond[39],

D carbon[40] and $C_{60}$ clathrate[30]. This strongly indicates that triple periodic Type-I, -II and -H carbon clathrates are lightweight mechanical engineering structure for important practical applications.

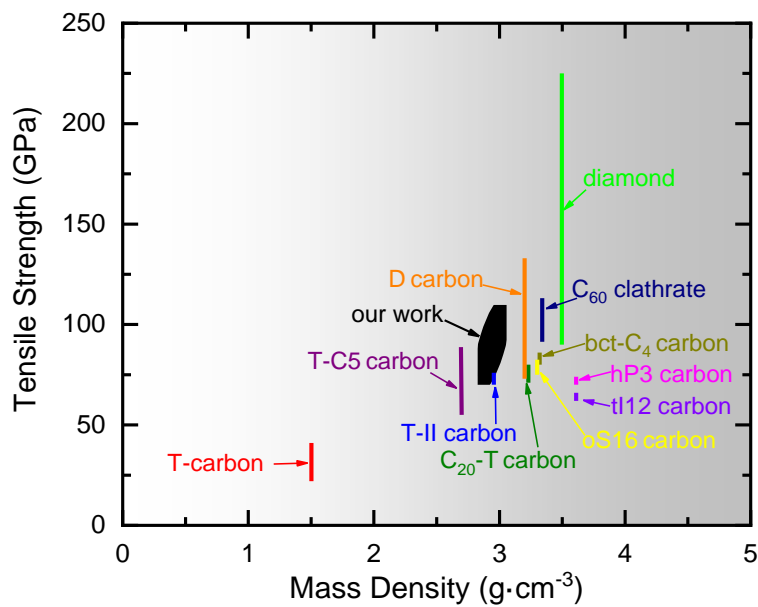

Figure 4 Tensile mechanical strength-mass density map of a variety of crystalline carbon allotropes[30, 37-42].

**3.4 Deformation Mechanisms and Critical Bond Configurations**

To reveal the deformation mechanisms underlying the unique mechanical responses of triple periodic Type-I, -II and -H carbon clathrates, the developments of their molecular structures subjected to specific uniaxial elongation and shearing loads are recorded. As an example, side-viewed snapshots of triple periodic Type-I carbon clathrate at critical strains under uniaxial tension and shearing are captured in Figure 5, where the $sp^3$-hybridized C-C bonds are colored on the basis of the values of bond length. As shown in Figure 5a, upon uniaxial tension, it is observed a non-uniform distribution in the elongation of $sp^3$-hybridized C-C bonds prior to critical strain of around 0.208. C-C Bonds that make small angles to the loading direction are the critical load-bearing configurations. For example, the C-C bonds that are parallel to the stretching direction are the most elongated configurations. As a consequence of non-uniform deformation of bond configurations and Poisson effect, all polyhedral cages are gradually misshaped during this elongation process. Intriguingly, as the strain approaches 0.208, localized bonds are broken as marked by the dash line. However, such strain-induced local bond dissociation does not result in sudden mechanical failure. As the strain is increased from 0.208 to 0.257, polyhedral cages become more misshaped because the major load-bearing shifts to the bond angles that is in the vicinity of the broken bonds, instead of further mechanical failure via bond breaking. The shift of load-bearing from bond to bond angles is mainly responsible for the negative tensile stiffness in the loading curve of Figure 2b. Once the elongation is over critical strain of around 0.257, highly deformed Type-I carbon clathrate catastrophically fails via brittle fractures at one cross-section, as indicated by snapshot at strain of 0.270, explaining the sudden deep drop of loading stress in Figure 2b. With regard to the case subjected to shearing load, triple periodic Type-I carbon clathrate presents similar characteristics of elastic deformation mechanisms to those under uniaxial tension, as indicated by Figure 4b. In contrast to the case under uniaxial tension, however, it shows significant plastic deformations as a result of global distribution of bond dissociations, explaining the re-rise in the loading stress after sudden deep drop of loading stress.

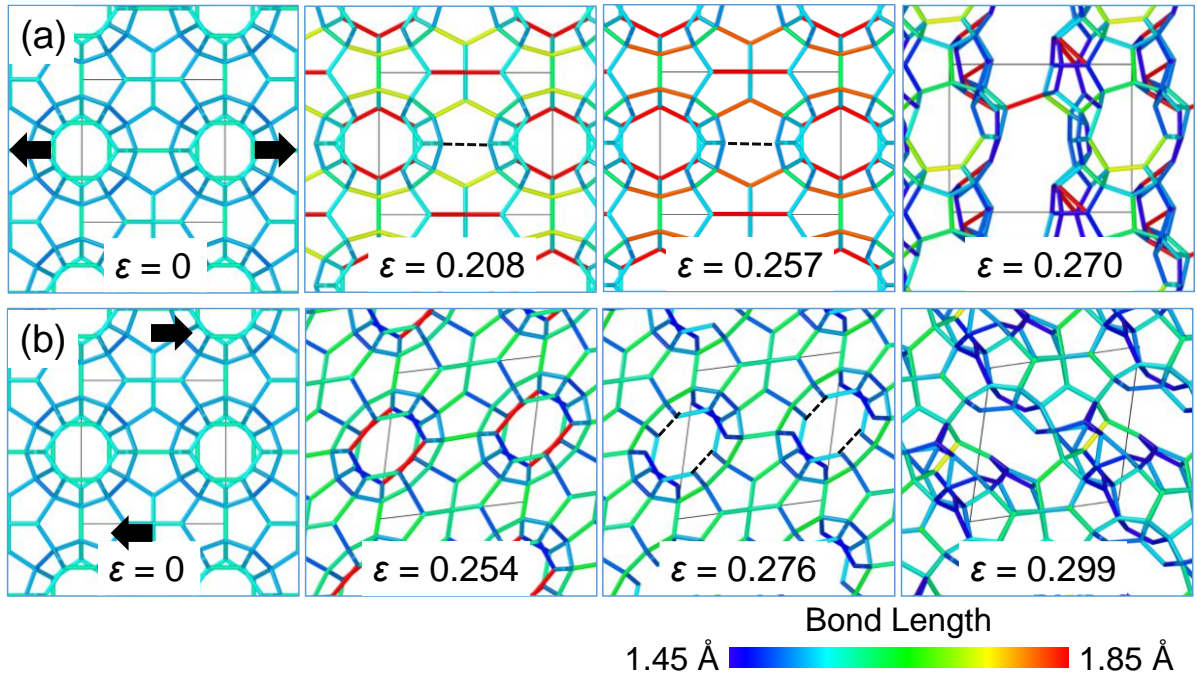

Figure 5 Mechanical deformation characteristics of triple periodic Type-I carbon clathrate. (a) and (b) Side-views of molecular structures of triple periodic Type-I carbon clathrate at critical strains subjected to <100> directional and (b) {001}<100> directional shearing loads, respectively. For clarification, the $sp^3$-hybridized C-C bonds are rendered based on the values of bond length. The black frames in the snapshots indicate the structure of unit cell of Type-I carbon clathrate.

Figure 6 shows the quantitative variations in the length of critical bonds and critical bond angles of triple periodic Type-I, -II and -H carbon clathrates with mechanical strain. The critical bonds and critical bond angles are defined by the bonds and bond angles that are the most load-bearing bond configurations under both uniaxial and shear strains, namely, the maximum increase in the bond length and bond angles. At zero strain (equilibrium state), Type-I, -II and -H carbon clathrates show average C-C bond lengths of around 1.555 Å, 1.548 Å and 1.553 Å, respectively, comparable to that of diamond (1.548 Å)[28]. Upon uniaxial tension, it is identified from Figures 6a and c that both critical bond length and bond angles of Type-I, -II and -H carbon clathrates are nonlinearly increased prior to the final deformation stage. Type-II and -H carbon clathrates show increase of the critical bond length from around 1.590 Å, 1.548 Å to 1.936 Å, 1.998 Å by 21.76%, 29.07%, respectively, indicating that critical bonds in Type-H carbon clathrate show higher deformability than those in Type-II one. However, the critical bond length in Type-I carbon clathrate is non-physically increased from 1.590 to 2.487 Å by around 56.42%. This is primarily attributed to the fact that, as the strain is over 0.208, the critical bonds have been broken and the corresponding atomic distance gradually increases because the surrounding bond angles of the critical bonds becomes the major load-bearing bond configuration, as indicated by Figure 5a. The following sudden rises in their critical bond lengths are indicative of catastrophically brittle fractures of the carbon clathrates. With regard to the critical bond angles, it is observed from Figure 6c that they are nonlinearly increased from about 111.90°, 119.95° and 120.04° to 126.21°, 134.33° and 136.05° by 12.79%, 11.99% and 13.34%, respectively. Obviously, subjected to uniaxial tension, the critical bonds of all carbon clathrates show much higher deformability than the critical bond angular configurations.

Upon shearing load, prior to the final deformational stage, both critical bonds and critical bond angles are monotonically increased with increasing shear strain. Note that the critical bonds and critical bond angles usually changes as the loading mode and loading direction are changed. From Figure 6c, the critical bonds in Type-I, -II and -H carbon clathrates are enlarged from around 1.571 Å, 1.548 Å and 1.567 Å to 1.891 Å, 1.818 Å and 1.637 Å by 20.37%, 17.44% and 4.47%, respectively, implying that they show distinct deformability. The elongation strains of critical bonds are revealed to be less than the applied global shear strains, indicating that the bond angular stretching also plays an important role in the global shear deformation. As seen from Figure 6d, the critical bond angles in Type-I, -II and -H carbon clathrates are enlarged from around 124.04°, 119.88°, 90.01° to 143.35°, 139.69° and 111.30° by 15.57%, 16.52% and 23.65%, respectively. The ranking of bond angular strain is opposite to that of bond strain. By comparison, the straining of critical bonds in Type-I carbon clathrate plays a more important role in the global shear deformation than that of critical bond angles, which is similar to the case of Type-I, -II and -H carbon clathrates subjected to uniaxial tension. As for Type-II carbon clathrate, however, both critical bond and critical angular stretching play comparable role in its global shear deformation. More intriguingly, the stretching of critical bond angles plays a dominant role in its global shear deformation. This clearly indicates that, upon shear load, although they are structurally dominated by polyhedral cages, the major load-bearing bond configurations in their structures are very different from each other.

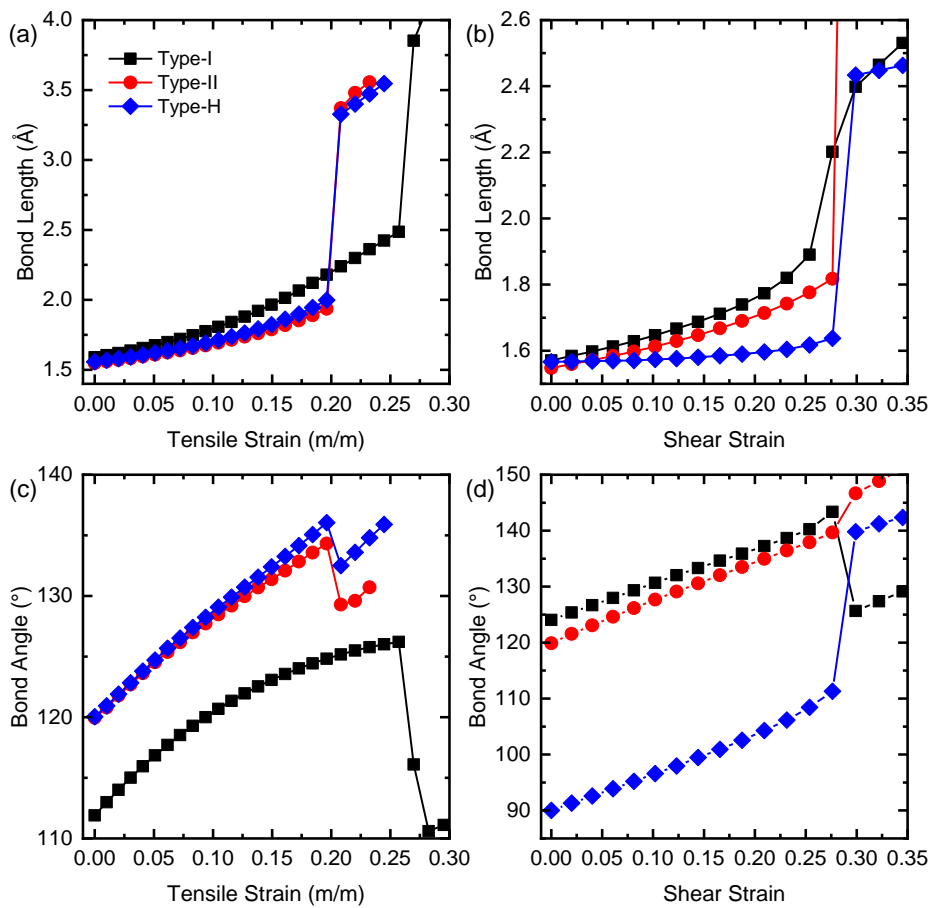

Figure 6 Quantitative analysis of critical bond configurations in carbon clathrates. Variations in the critical bond lengths of triple periodic Type-I, -II and -H carbon clathrates with (a) <100> tensile strain

and (b) {001}<100> shear strain. Changes in the critical bond angles of triple periodic Type-I, -II and -H carbon clathrates under (c) <100> directional elongation and (d) {001}<100> shear load.

**3.5 Strain-engineered Modulation in the Electronic Properties of Carbon Clathrates**

As is known, mechanical strain is an effective way to tune the physical properties of crystalline materials, including electronic, transport, and optical properties, because mechanical strain causes structural changes in crystals. Here, the electrical conductivities of mechanically deformed Type-I, -II and -H carbon clathrates are accordingly explored. As an example, Figure 7 shows the representative electronic band structures of Type-I carbon clathrate subjected to different strains along <100> direction. Obviously, the electronic band structures of Type-I carbon clathrate are greatly dictated by tensile strain. With increasing the elastic strain, the conduction band minimum (CBM) tends to gradually move to the Fermi level. Once the tensile strain is applied to 0.270, the band states start to cross the Fermi level. It is worth noting that the position of the CBM is relatively steep, namely, the second derivative of the energy band with respect to $k$ is relatively large. The conductive charges are concentrated on the covalent bonds of carbon atoms that are greatly changed as a result of application of global tensile strain. As a result of weakening of carbon covalent bonds, the conductive charges are released, thereby causing the deformed structure more conductive. This strongly indicates that Type-I carbon clathrate shows low effective mass and good carrier migration properties. Such large tunability in the electronic properties has been also identified in diamond under compressive shear strain[43].

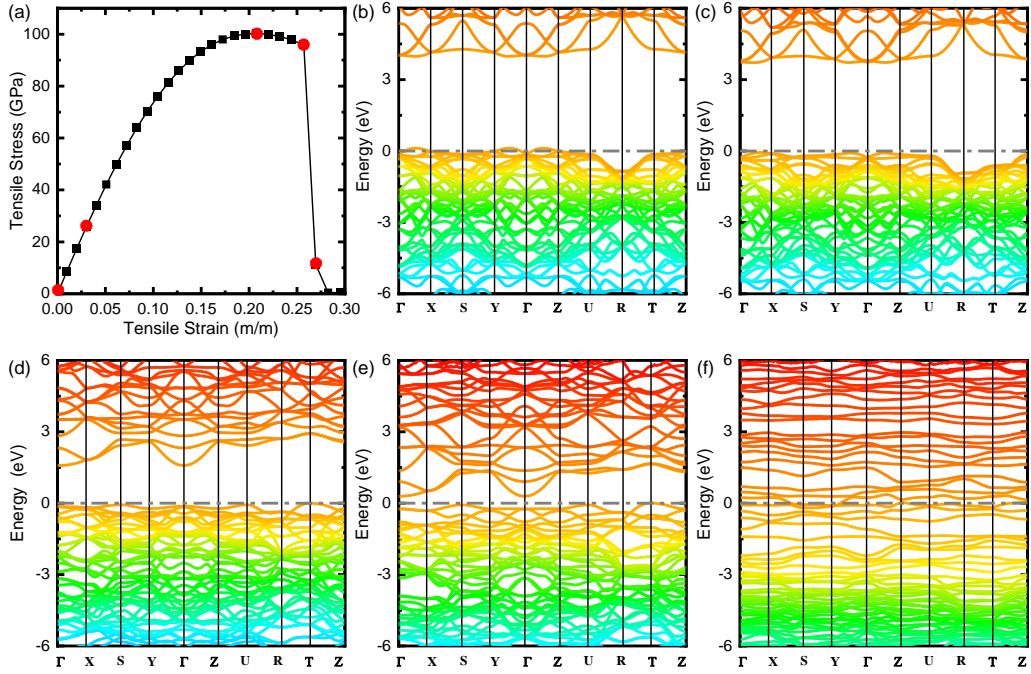

Figure 7 Electronic properties of mechanically deformed Type-I carbon clathrate. (a) Tensile stress-strain curve of triple periodic Type-I carbon clathrate subjected to <100> directional load. (b)-(f) Electronic band structures of triple periodic Type-I carbon clathrate at different strains that are red-circle marked in the loading curve.

Figure 8 further collects the band gaps of Type-I, -II and -H carbon clathrates subjected to different uniaxial tensile and shear strains. As is seen, under zero strain (equilibrium state), the band gaps of Type-I, -II and -H carbon clathrates are determined to be around 3.86 eV, 3.74 eV and 3.35 eV, respectively, indicating that they are intrinsically indirect band gap semiconductors. Upon both uniaxial tensile and shear strain, bond gaps of those three carbon clathrates are nonlinearly reduced with increasing strain, and the reduction tendency in band gaps becomes more pronounced as the strain is enlarged. By comparison, upon uniaxial tension below 0.161, in terms of the reduction of band gaps with a given strain, they can be sorted as Type-I > Type-II > Type-H carbon clathrates. Intriguingly, as the uniaxial strain approaches around 0.161, Type-I, -II and -H carbon clathrates yield identical band gap of around 2.38 eV. Upon shear straining, it is also identified that the reduction in the band gaps of the three carbon clathrates becomes more significant with increasing elastic shear strain, and similar band gaps can be achieved as the three carbon clathrates are highly sheared. Prior to mechanical failure, the three deformed carbon clathrates show band gap of 1.5-2.0 eV, depending on the clathrate structure. Once they mechanically fail, their band gaps suddenly drop to almost zero, indicating that they become conducting structures. Such significant changes in band gaps of those carbon clathrates with both uniaxial tensile and shear strains enable them as potential electromechanical devices.

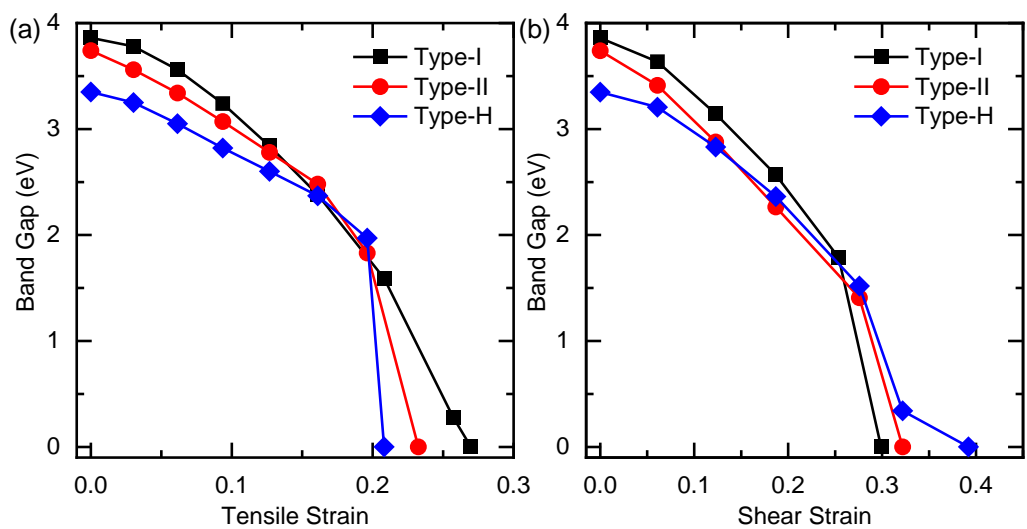

Figure 8 Variations in the band gap of Type-I, -II and -H carbon clathrates with (a) uniaxial strain of <100> direction and (b) shear strain of {001}<100> slip system, respectively.

4. **Conclusions**

Clathrates are $sp^3$-bonded cage-like 3D structures that have drawn remarkable attention in condensed-matter physics and chemistry due to their unique mechanical and electronic properties. The mechanical stability and mechanical anisotropy of hypothetical Type-I, Type-II, and Type-H clathrates comprised of different carbon frameworks subjected to uniaxial tensile and shear loads are examined by first-principles calculations. It is demonstrated that the three carbon clathrates show unique tensile and shear mechanical properties that greatly vary with carbon framework and crystallographic orientation. Upon uniaxial tension, both Type-II and -H carbon clathrates exhibit similar anisotropy with a hexagonal star-shaped pattern in the ideal tensile strength, while Type-I carbon clathrate show a dodecagon-shaped anisotropic pattern in the ideal tensile strength. It is identified a quadrangle-shaped pattern in the anisotropy of tensile Young's modulus of Type-I carbon clathrate; however, Type-II and -H carbon clathrates show quite isotropic tensile Young's modulus. Under shear load, Type-I, II and -H carbon clathrates show quadrangle-shaped, hexagon-shaped and gear-like patterns in the anisotropy of ideal shear strength, respectively. Intriguingly, the three carbon clathrates present isotropic shear Young's modulus, although they are different carbon framework structures. By comparison, those carbon clathrates yield better mechanical performance of ideal tensile strength than that of lightweight T-carbon and heavyweight bct-C4, hP3, tI12, oS16, T-II carbon allotropes, as well as comparable to heavyweight diamond, D carbon and $C_{60}$ clathrate, indicating that they are lightweight mechanical robust 3D metastructures for important applications. Furthermore, the three carbon clathrates are intrinsically indirect band gap semiconductors, and their band gaps structures can be greatly reduced by both tensile and shear strains.

**Acknowledgments**

This work is financially supported by the National Natural Science Foundation of China (Grant Nos. 11772278, 11904300 and 11502221), the Jiangxi Provincial Outstanding Young Talents Program (Grant No. 20192BCBL23029), the Fundamental Research Funds for the Central Universities (Xiamen University: Grant Nos. 20720180014 and 20720180018). Y. Yu and Z. Xu from Information and Network Center of Xiamen University for the help with the high-performance computer.

**Competing interests**

The authors declare that they have no competing interests.

# References

Primary Sources

Secondary Sources

Uncategorized References

1.   Bradley, D.K., et al., *Diamond at 800 GPa.* Phys Rev Lett, 2009. **102**(7): p. 075503.

2.   Kroto., H.W., et al., *C60: buckministerfullerene.* Nature, 1985. **318**.

3.   Iijima., S., *Helical microtubules of graphitic carbon.* Nature, 1991. **354**.

4.   Cook, M.W. and P.K. Bossom, *Trends and recent developments in the material manufacture and cutting tool application of polycrystalline diamond and polycrystalline cubic boron nitride.* International Journal Of Refractory Metals & Hard Materials, 2000. **18**(2-3): p. 147-152.

5.   Thostenson, E.T., Z.F. Ren, and T.W. Chou, *Advances in the science and technology of carbon nanotubes and their composites: a review.* Composites Science And Technology, 2001. **61**(13): p. 1899-1912.

6.   Novoselov, K.S., et al., *Electric field effect in atomically thin carbon films.* Science, 2004. **306**(5696): p. 666-9.

7.   Perreault, F., A. Fonseca de Faria, and M. Elimelech, *Environmental applications of graphene-based nanomaterials.* Chem Soc Rev, 2015. **44**(16): p. 5861-96.

8.   Kasper, J.S., et al., *Clathrate Structure of Silicon Na8Si46 and NaxSi136 (x < 11).* Science, 1965. **150**(3704): p. 1713-4.

9.   Herrmann, R.F.W., et al., *Electronic structure of Si and Ge gold-doped clathrates.* Physical Review B, 1999. **60**(19): p. 13245-13248.

10. Guloy, A.M., et al., *A guest-free germanium clathrate.* Nature, 2006. **443**(7109): p. 320-323.

11. Benedek, G., et al., *Hollow Diamonds - Stability And Elastic Properties.* Chemical Physics Letters, 1995. **244**(5-6): p. 339-344.

12. Munetoh, S., et al., *Molecular-dynamics studies on solid phase epitaxy of guest-free silicon clathrates.* Thermoelectric Materials 2001-Research And Applications, 2001. **691**: p. 457-462.

13. Reny, E., et al., *A Na-23 NMR study of NaxSi136 and Na8Si46 silicon clathrates.* Comptes Rendus De L Academie Des Sciences Serie Ii Fascicule C-Chimie, 1998. **1**(2): p. 129-136.

14. Zhang, J., et al., *Pseudo-topotactic conversion of carbon nanotubes to T-carbon nanowires under picosecond laser irradiation in methanol.* Nat Commun, 2017. **8**(1): p. 683.

15. Tanigaki, K., et al., *Silicon and germanium clathrates with magnetic elements.* Nanonetwork Materials: Fullerenes, Nanotubes And Related Systems, 2001. **590**: p. 495-498.

16. Reny, E., et al., *Structural characterisations of the NaxSi136 and Na8Si46 silicon clathrates using the Rietveld method.* Journal Of Materials Chemistry, 1998. **8**(12): p. 2839-2844.

17. Bobev, S. and S.C. Sevov, *Synthesis and characterization of stable stoichiometric clathrates of silicon and germanium: Cs8Na16Si136 and Cs8Na16Ge136.* Journal Of the American Chemical Society, 1999. **121**(15): p. 3795-3796.

18. Ramachandran, G.K., et al., *Synthesis and X-ray characterization of silicon clathrates.* Journal Of Solid State Chemistry, 1999. **145**(2): p. 716-730.

19. Sheng, X.L., et al., *T-carbon: a novel carbon allotrope.* Phys Rev Lett, 2011. **106**(15): p. 155703.

20. Blase, X., *Quasiparticle band structure and screening in silicon and carbon clathrates.* Physical Review B, 2003. **67**(3).

21. Blase, X., et al., *Exceptional ideal strength of carbon clathrates.* Phys Rev Lett, 2004. **92**(21): p. 215505.

22. Connetable, D. and X. Blase, *Electronic and superconducting properties of silicon and carbon clathrates.* Applied Surface Science, 2004. **226**(1-3): p. 289-297.

23. Zipoli, F., M. Bernasconi, and G. Benedek, *Electron-phonon coupling in halogen-doped carbon clathrates from first principles.* Physical Review B, 2006. **74**(20).

24. Rey, N., et al., *First-principles study of lithium-doped carbon clathrates under pressure.* Journal of Physics: Condensed Matter, 2008. **20**(21): p. 215218.

25. Connetable, D., *First-principles calculations of carbon clathrates: Comparison to silicon and germanium clathrates.* Physical Review B, 2010. **82**(7).

26. Wang, J.-T., et al., *Phase stability of carbon clathrates at high pressure.* Journal of Applied Physics, 2010. **107**(6).

27. Colonna, F., A. Fasolino, and E.J. Meijer, *Structure and thermodynamic stability of carbon clathrates: A Monte Carlo study.* Solid State Communications, 2012. **152**(3): p. 180-184.

28. Zeng, T., et al., *Li-Filled, B-Substituted Carbon Clathrates.* J Am Chem Soc, 2015. **137**(39): p. 12639-52.

29. Kolezynski, A. and W. Szczypka, *First-Principles Study of the Electronic Structure and Bonding Properties of X8C46 and X8B6C40 (X: Li, Na, Mg, Ca) Carbon Clathrates.* Journal Of Electronic Materials, 2016. **45**(3): p. 1336-1345.

30. Li, Z., et al., *Superhard superstrong carbon clathrate.* Carbon, 2016. **105**: p. 151-155.

31. Kresse, G. and J. Furthmuller, *Efficient iterative schemes for ab initio total-energy calculations using a plane-wave basis set.* Physical Review B, 1996. **54**(16): p. 11169-11186.

32. Kresse, G. and D. Joubert, *From ultrasoft pseudopotentials to the projector augmented-wave method.* Physical Review B, 1999. **59**(3): p. 1758-1775.

33. Blochl, P.E., *Projector Augmented-Wave Method.* Physical Review B, 1994. **50**(24): p. 17953-17979.

34. Perdew, J.P., K. Burke, and M. Ernzerhof, *Generalized gradient approximation made simple.* Physical Review Letters, 1996. **77**(18): p. 3865-3868.

35. Setyawan, W. and S. Curtarolo, *High-throughput electronic band structure calculations: Challenges and tools.* Computational Materials Science, 2010. **49**(2): p. 299-312.

36. Roundy, D. and M.L. Cohen, *Ideal strength of diamond, Si, and Ge.* Physical Review B, 2001. **64**(21).

37. Zhang, S.H., et al., *High-throughput screening for superhard carbon and boron nitride allotropes with superior stiffness and strength.* Carbon, 2018. **137**: p. 156-164.

38. Li, D., et al., *Modulated T carbon-like carbon allotropes: an ab initio study.* RSC Advances, 2014. **4**(33): p. 17364.

39. Telling, R.H., et al., *Theoretical strength and cleavage of diamond.* Phys Rev Lett, 2000. **84**(22): p. 5160-3.

40. Fan, D., et al., *D-carbon: Ab initio study of a novel carbon allotrope.* J Chem Phys, 2018. **149**(11): p. 114702.

41. Wang, J.Q., et al., *C 20 - T carbon: a novel superhard sp (3) carbon allotrope with large cavities.* J Phys Condens Matter, 2016. **28**(47): p. 475402.

42. Pang, D.-D., et al., *Properties of a predicted tetragonal carbon allotrope: First principles study.* Diamond and Related Materials, 2018. **82**: p. 50-55.

43. Liu, C., et al., *Superconductivity in Compression-Shear Deformed Diamond.* Phys Rev Lett, 2020. **124**(14): p. 147001.